\documentclass{emulateapj}
\usepackage{float, amsmath,verbatim}
\pdfoutput=1

\shorttitle{Ion Acceleration in Relativistic Shocks} 
\shortauthors{Martins}

\begin{document}

\title{Ion dynamics and acceleration in relativistic shocks} 

\author{S. F. Martins\altaffilmark{1}}
\author{R. A. Fonseca\altaffilmark{1,2}}
\author{L. O. Silva\altaffilmark{1}}
\author{W. B. Mori\altaffilmark{3}}
\altaffiltext{1} {GoLP/Instituto de Plasmas e Fus\~ao Nuclear, Instituto Superior T\'ecnico, Lisbon, Portugal: samuel.martins@ist.utl.pt}
\altaffiltext{2} {DCTI/Instituto Superior de Ci\^encias do Trabalho e da Empresa, Lisbon, Portugal}
\altaffiltext{3} {University of California Los Angeles, Los Angeles, USA}
\subjectheadings{acceleration of particles -- collisionless shocks -- gamma rays: bursts }

\begin{abstract}

Ab-initio numerical study of collisionless shocks in electron-ion unmagnetized plasmas is performed with fully relativistic particle in cell simulations. The main properties of the shock are shown, focusing on the implications for particle acceleration. Results from previous works with a distinct numerical framework are recovered, including the shock structure and the overall acceleration features. Particle tracking is then used to analyze in detail the particle dynamics and the acceleration process. We observe an energy growth in time that can be reproduced by a Fermi-like mechanism with a reduced number of scatterings, in which the time between collisions increases as the particle gains energy, and the average acceleration efficiency is not ideal. The in depth analysis of the underlying physics is relevant to understand the generation of high energy cosmic rays, the impact on the astrophysical shock dynamics, and the consequent emission of radiation.

\end{abstract}

\section{Introduction}

The dynamics of relativistic particle acceleration in collisionless shocks is of great importance to several astrophysical scenarios. The acceleration of electrons, positrons, or ions in various structures such as active galactic nuclei, gamma-ray bursts, pulsar wind nebulae, and supernova remnants results in energetic particles that can then scatter in magnetic and electric fields and emit synchrotron radiation (see, for instance, Jones \& Ellison 1991). Despite their relevance to understanding the radiation collected in astronomical observations, the underlying processes inherent to the acceleration are not yet fully understood. In this context, full kinetic simulations play an important role in the assessment of particular physical mechanisms relevant for astrophysical shocks, namely in the study of the nonlinear growth of the Weibel instability \citep{weibel59, medvedev99} with magnetic field generation \citep{silva03, fonseca03, fred04, nish05}, and particle acceleration in the percursor region \citep{silva06}. Nevertheless, the self-consistent modeling of a relativistic collisionless shock, from first principles, is computationally very demanding, as large temporal and spatial scales push the numerical techniques to conditions still unexplored. Hence, full kinetic simulations require massive computational resources, optimized algorithms, methods for improved energy conservation, and also advanced visualization diagnostics.

Recently, progress was made in the understanding of electron-ion shock formation and electron-positron acceleration in relativistic unmagnetized shocks \citep{spitk08a,spitk08b}. These studies with particle in cell (PIC) simulations confirmed the capability of these structures to effectively accelerate electrons, which is identified by the development of a non-thermal tail in the energy spectrum. Similar approaches with large-scale self-consistent modelling can provide valuable input to improve Monte Carlo methods (e.g., Bednarz \& Ostrowski 1998,  Ellison \& Double 2004), and to support the development of analytical models (e.g., Kirk et al. 2000, Achterberg et al. 2001, Keshet 2006).    

Here, we examine a relativistic electron-ion unmagnetized shock with ab-initio relativistic PIC simulations, and apply a full particle tracking diagnostic to better understand the particle dynamics and the acceleration mechanism. We then show that non-thermal particle acceleration occurs through a small number of scatterings in the shock front, in a Fermi-like process ($\Delta E = \alpha E$) with increasing time between each energy gain. Our simulations and initial data analysis follow and confirm previous results obtained with a different PIC framework \citep{spitk08a,spitk08b}, which reveals the robustness of the numerical results with distinct algorithms and implementations. In \S2, we present the simulation results of the electron-ion plasma shock with a reduced ion to electron mass ratio of 32. In \S3, we leverage on the OSIRIS \citep{fonseca02} particle tracking and data processing tools to analyze the particle dynamics and their acceleration process, by focusing on the time evolution of the main physical quantities of the most energetic particles. A discussion of the acceleration mechanism and the conclusions are presented in \S4.

\section{Shock formation and evolution} \label{secform}

Numerical simulations were performed with OSIRIS, a fully relativistic, electromagnetic, and massivelly parallel PIC code which has been used in many different physical scenarios, such as astrophysics (e.g., Silva et al. 2003), laser/plasma accelerators (e.g., Mangles et al. 2004, Tsung et al. 2004), nanoplasma dynamics (e.g., Peano et al. 2006), and fast-ignition (e.g., Ren et al. 2004).

We simulate a two-dimensional system of a cold unmagnetized electron-ion plasma with mass ratio $m_i/m_e = 32$ ($m_i$ and $m_e$ the ion and electron mass, respectively), and evolve it until evidence of a non-thermal acceleration tail in the downstream particle spectrum is achieved. To generate the shock, the plasma stream is launched from the right wall with proper velocity $u=\gamma \beta = -20$, and minimal thermal dispersion from randomized particle injection. This neutral plasma stream is reflected from a rigid boundary at the left wall (this is one of the most direct methods to generate shocks in simulations; see, for instance, Forslund et al. 1970, Jones \& Ellison 1991, or Spitkovsky 2008a,b). The computational domain is $50~c/\omega_p$ in the transverse direction and $280~c/\omega_p$ in the longitudinal direction, with $ c/\omega_p = (4\pi e^2n/\gamma m_i c^2)^{-1/2}$ the ion skin depth for a number density $n$ and relativistic ion mass $\gamma m_i$; $e$ is the elementary charge and $c$ the speed of light in vacuum. A time step of $0.012/\omega_p$ is used. The system is numerically resolved with 10 cells per electron skin depth in both directions, thus ensuring that the dynamics of the lighter species is accurately modeled. Taking advantage of second order particle shapes \citep{esir01} and current smoothing compensation, we use 2 particles per cell (ppc) for each species, which is equivalent to 6 ppc in the shocked gas, for which the energy conservation is equivalent to 16 ppc with linear particle shapes \citep{fonseca08}. Lower mass ratios, higher grid resolutions, and more particles per cell were tested, showing an overall qualitative and quantitative result convergence. 

\begin{figure} 
\centering
\plotone{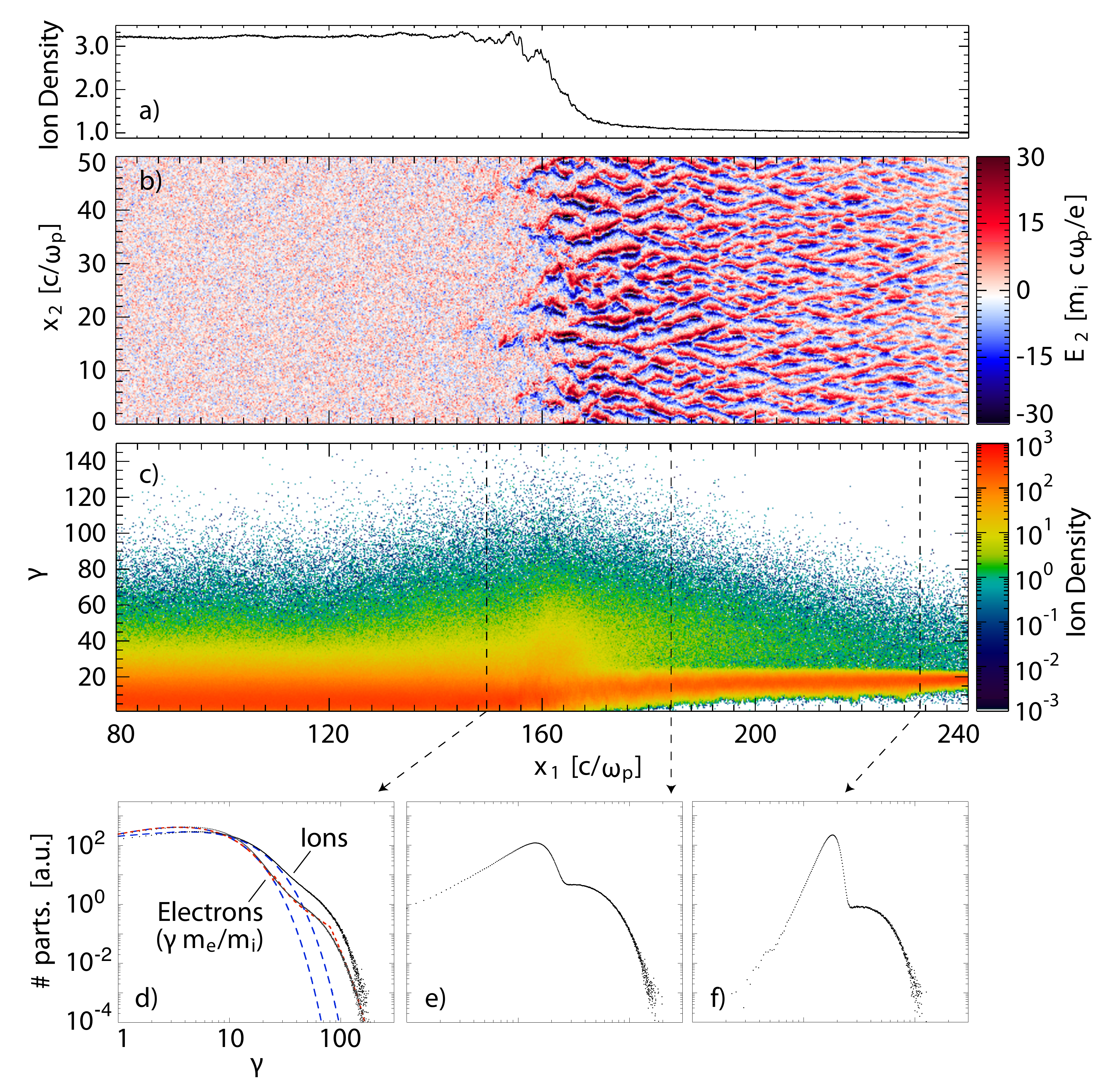}
\caption{Steady state structure of a $m_i/m_e=32$ collisionless shock at $t = 360/\omega_p$ after first counter-streaming interaction. a) Transverse average of the ion density; b) Transverse electric field; c) Ion energy spectrum along simulation box; d-f) Ion spectra for the downstream, the shock front, and the upstream regions, respectively. The downstream inset includes also the electron spectrum (scaled down by the mass ratio $m_e/m_i$) with relativistic Maxwellian fits for both species - blue dashed lines -, and a fit to the electron spectrum of a relativistic Maxwellian with a power law trimmed by an exponential cut-off for higher energies - red dotted line.}
\label{fig:shocksummary}
\end{figure}

The main physical processes that generate the shock are observed when the reflected particle stream interacts with the incoming plasma, which leads to the growth of the Weibel instability, particle thermalization, and the generation of electric and magnetic turbulence. The turbulence slows down the flow, which generates the shock as a density compression that propagates in the positive $x_1$ direction (Fig.~\ref{fig:shocksummary}a). Simulation results agree with the hydrodynamical jump conditions \citep{blandford76, spitk08a}; the steady state velocity of the shock, $\beta_{\mathrm{shock}} \simeq 0.48$, and the corresponding density compression obtained directly by particle number conservation, $n_2/n_1 = 3.1$. We emphasize that, given the simulation configuration, all quantities discussed are measured in the downstream frame.

Fig.~\ref{fig:shocksummary} also shows other relevant physical quantities after the shock has achieved a steady state. Of relevance to the acceleration process, transverse electric fields (Fig.~\ref{fig:shocksummary}b) arise in the linear stage of the Weibel instability, associated with space charge effects, as the two counter-propagating plasma streams pinch/filament with different rates because of their different temperatures \citep{tzouf06}. The spatial symmetries of this field have direct impact on the overall transverse momentum acquired by the particles (c.f. \S\ref{sec:accel}). In addition, the energy deposited in the magnetic field reaches $\epsilon_B \equiv B^2/4\pi \gamma n m c^2 \simeq 15-20\%$, similarly to pair shocks \citep{spitk08b}, and to the $m_i/m_e = 16$ case. This observation, coupled with the average value of $\mathbf v_{\mathrm{drift}}= \mathbf E \times \mathbf{B}/B^2$ across the shock front associated with the structure of the self-consistent fields, suggests also the origin of the particle trapping mechanism in the shock front for both positrons/ions and electrons (Martins et al., in preparation), similar to what is observed in Earth bow shocks \citep{burgess07}.

In accordance with results obtained by Spitkovsky (2008b), the energy spectrum of the ions (Fig.~\ref{fig:shocksummary}c-f) is significantly different across the longitudinal direction. The upstream region is dominated by the quasi-monoenergetic negative flow of particles, and contains a lower density returning stream of heated particles that already escaped the shock region (or were never trapped). Despite the strong thermalization of the shocked gas, evidence of non-thermal particle acceleration can be observed in the downstream ion spectrum (Fig.~\ref{fig:shocksummary}d), where a fit to a pure relativistic Maxwellian does not account for the high-energy tail. The non-thermal spectrum of both electrons and ions can be fitted with a power law ($\gamma^{-p}$, with $p = 2.3-2.6$) and an exponential cut-off defined by $\exp[-(\gamma-\gamma_{\mathrm{cut}})/\Delta \gamma_{\mathrm{cut}}]$, with $\gamma_{\mathrm{cut}} = 80$ and $\Delta \gamma_{\mathrm{cut}}=15$ for the ion species. Electrons reach energies that are higher by the mass ratio $m_i/m_e$, thus spanning more than three orders of magnitude in energy. The high-energy tail (for $\gamma > 40$) has $\sim$2\% of the total number of ions in the downstream slice analyzed, and accounts for $\sim$10\% of the total ion energy in that spatial region. This confirms previous results obtained for a pair plasma configuration by \cite{spitk08b}, leading to the conclusion that the generated spectra and acceleration efficiency are not very sensitive to the mass ratio of the species, at least in two-dimensional simulations. Finally, additional propagation time leads to a linear increase of the non-thermal tail span to higher energies, and results indicate a fit with the same power law index. 

\section{Particle dynamics \& acceleration} \label{sec:accel}

For a detailed analysis of the particle dynamics and acceleration mechanism, the OSIRIS particle tracking features  \citep{fonseca08} were used to follow the trajectories of the most energetic particles, selected in a first scanning simulation.

\begin{figure} 
\centering
\plotone{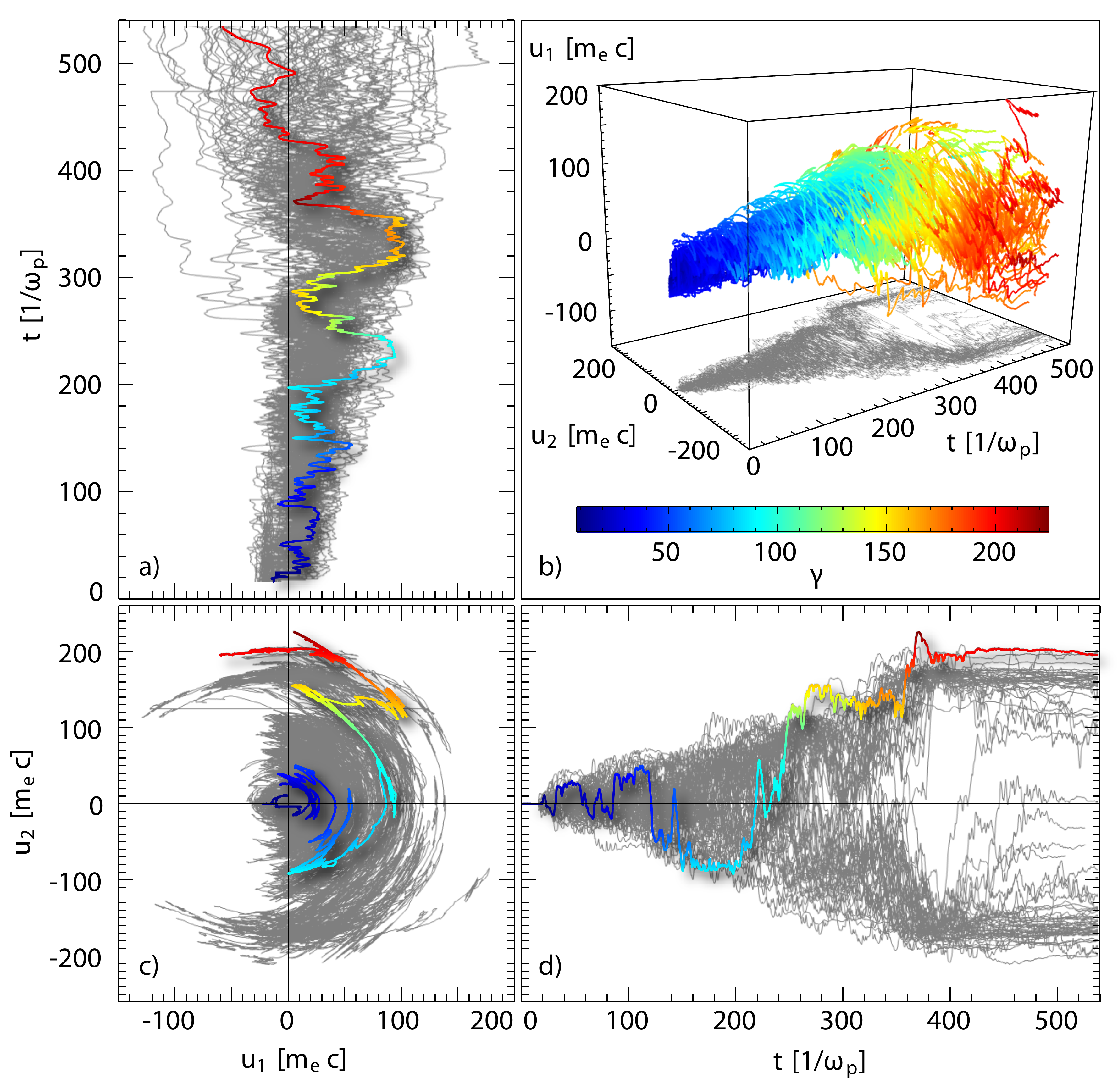}
\caption{Momentum-time trajectories of the 80 most energetic particles, chosen at $t = 360/\omega_p$ and evolved until $t = 550/\omega_p$. The trajectory in color corresponds to the most energetic particle. a) Longitudinal momentum time evolution ($u_1\times t$), b) Longitudinal/transverse momentum time evolution ($u_1 \times u_2 \times t$), c) Longitudinal/transverse momentum evolution ($u_2\times u_1$), d) Transverse momentum time evolution ($u_2\times t$).}
\label{fig:p1p2t}
\end{figure}

Acceleration occurs for both longitudinal and transverse momentum, but with different dynamics in each direction (Fig.~\ref{fig:p1p2t}). On one hand, the approximately null space average of the transverse electric field (Fig.~\ref{fig:shocksummary}b) leads to a symmetric acceleration of the particles in the $x_2$ direction (Fig.~\ref{fig:p1p2t}c-d). On the other hand, the longitudinal momentum shows an average increase over time, and the particles reach farther into the upstream region, until escaping (Fig.~\ref{fig:p1p2t}a) mostly to the downstream region. The propagation of the particles encompasses several magnetic rotations in which momentum is transferred between directions, leading to the oscillations observed in $u_1\times t$ and $u_2\times t$. The ions are able to remain inside the accelerating structure by performing long drifts in the transverse direction ($u_2 \gg u_1$). For the most energetic particle (colored trajectory), the final energy gain occurs with high angle variation from the transverse to the longitudinal direction ($u_2$ nearly constant in Fig.~\ref{fig:p1p2t}c-d); the particle is then reflected from the upstream, and escapes the shock region into the downstream plasma (final $u_1 < 0$). After escaping, the particle sustains a large transverse momentum, and performs long drifts in the transverse direction with constant energy. Also, it should be emphasized that the behavior observed for this particle is representative of the accelerated particles. 

\begin{figure} 
\centering
\plotone{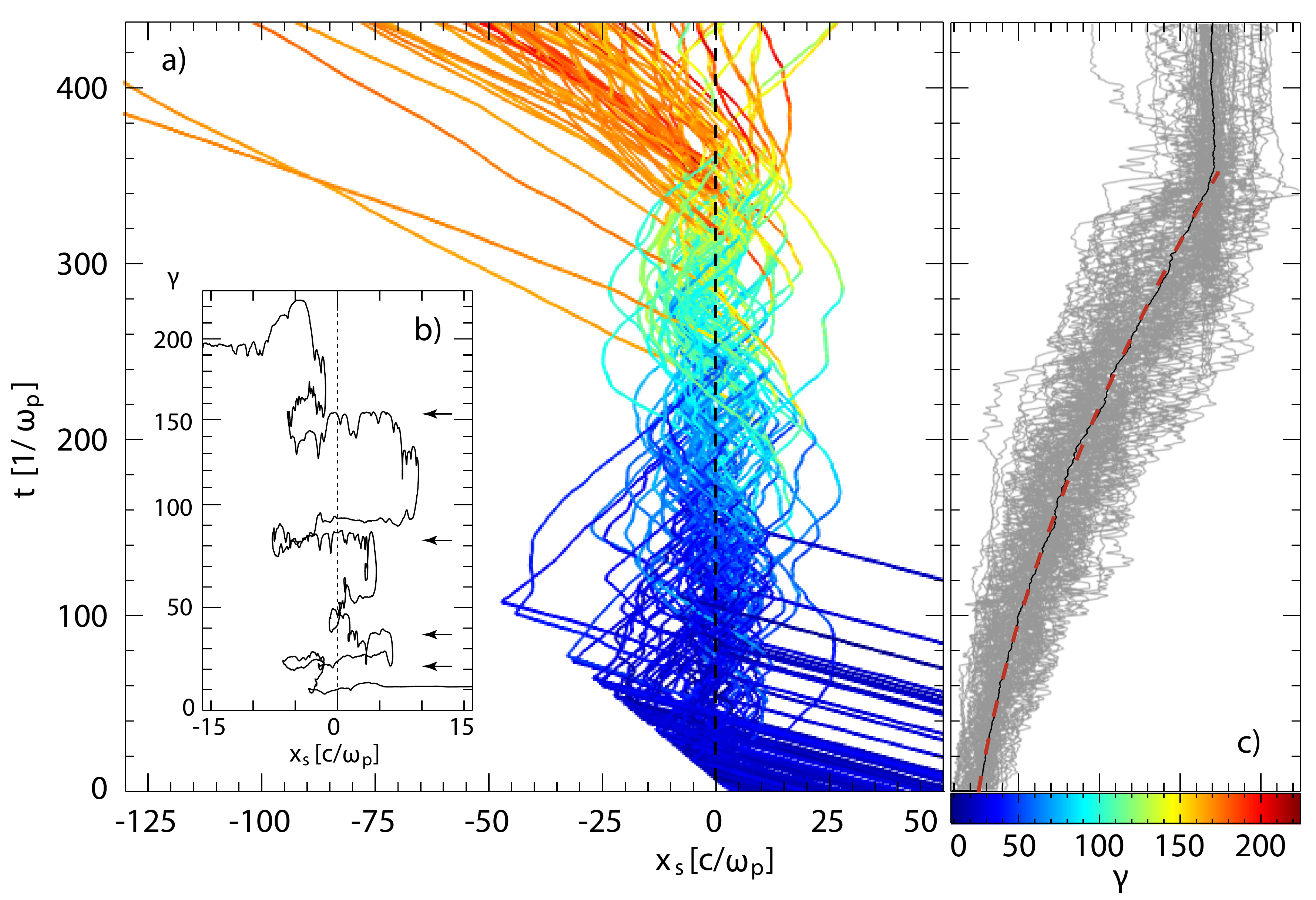}
\caption{Ion dynamics and energy evolution in time. a) Longitudinal ion position relative to the shock front ($x_s \equiv x_1 - x_{\mathrm{shock}}$) as a function of time (color in trajectories associated with energy) for the 80 most energetic particles (chosen at $t = 360/\omega_p$), b) Energy as a function of the position relative to the shock front for the most energetic particle; arrows indicate energy gains, c) Energy evolution in time for the same 80 ions (gray) with $t=0$ the trapping moment of each particle, corresponding average (solid line), and fit of $E(t) \simeq E_0 (1+\alpha)^{t/(\tau_0 + st)}$ with $\alpha = 0.81$, $s=0.11$, and $\tau_0 = 50.6/\omega_p$ (dashed line). The final flattening occurs as particles escape the shock, and leave the accelerating region.}
\label{fig:angle}
\end{figure}

The energy gain and the interaction with the shock region are depicted in Fig.~\ref{fig:angle}a, which shows the time evolution of the longitudinal position relative to the shock front for the 80 most energetic ions. Identically to the particle motion in pair plasmas observed in Spitkovsky 2008b, ions gain energy after being trapped as they perform multiple oscillations in the shock region until they finally escape, mainly to the downstream region. The wall reflections observed in Fig.~\ref{fig:angle}a are close to the injection point only at the beginning of the simulation, and do not affect the overall process. After the shock is formed, particles from the unshocked gas can be directly trapped, without reaching the downstream. The particle gains energy from the electric fields of the upstream, and then crosses the shock region until being reflected in the downstream (Fig.~\ref{fig:angle}b). We emphasize that, since the simulation is performed in the downstream frame, no significant acceleration occurs when particles are reflected on the downstream shocked gas. Accelerations occur rapidly and $\Delta E \simeq E$ is typically observed (Fig.~\ref{fig:angle}b), as expected in a Fermi process (Fermi 1949). The maximum energy reached is $\gamma_{\mathrm{final}} \simeq 170$. Since for a single bounce $\Delta E \simeq E$, and the initial ion energy is $\gamma_{\mathrm{initial}} \simeq 18$, that final energy can only be achieved after several shock crossings. It is important to notice that, unlike a Fermi mechanism, the final energy is usually obtained after only 3-5 effective collisions in a continuously evolving shock.

In the standard formalism of Fermi acceleration, the energy evolution is written as $E(N) = E_0 \exp(N)$, with $E_0$ the initial energy and $N$ the number of energy gains (assumed very large: $N \gg 1$), and its time dependence can therefore be estimated by relating $N$ with time $t$. The initial estimate by Fermi assumed a constant time between energy gains $\tau_{\mathrm{coll}}=\tau_0$, thereby leading to an exponential growth of the energy with time: $E(t) = E_0 \exp(t/\tau_0)$. To account for the small number of discrete energy gains observed in the simulation, we now write $E(N) = E_0(1+\alpha)^N$, where we include the constant fractional energy gain $\alpha \le 1$ in the form $\Delta E = \alpha E$, already considered by Fermi. The value $\alpha$ was estimated with an individual analysis of the energy gains of each tracked particle. Fitting the data of 52 of the 80 trajectories to $(1+\alpha)^N$ yields $\alpha \simeq 0.81$ (particular energy fluctuations of the remaining particles did not allow for a clear identification of all scatterings, defined as $x_1$ propagation direction inversion with $>50\%$ energy gain - see arrows in Fig.~\ref{fig:angle}b). 

Further analysis of several particle trajectories in the simulation shows an effect that decreases the energy growth with time, namely the increase of time between energy gains as the particle accelerates and the shock structure evolves. A simple model can be obtained using $\tau_{\mathrm{coll}}=\tau_0 + s t$, for $\tau_0$ the initial collision time, and where $s$ is the rate of change of the time between collisions. We thus get an approximate fit $N(t) = t/(\tau_0 + s t)$, and the energy evolution in time becomes $E(t) \simeq E_0 (1+\alpha)^{t/(\tau_0 + st)}$. We emphasize that this expression is only valid for a limited time $t$, and thus implicitly assumes particles escape the accelerating region with a maximum of $N(t \to \infty) = 1/s$ collisions. Fig.~\ref{fig:angle}c presents a fit to $E(t)$ with $s=0.11$ and $\tau_0=50.6/\omega_p$. Alternatively, the collision time can be written as a function of energy with $\tau_{\mathrm{coll}} = \tau_0[E(t)/E_0]^d$ (an energy dependence of $\tau_{\mathrm{coll}}$ is also observed in other scenarios, as in the Earth's bow shock acceleration, Kis et al. 2004). For this case, a numerical fit yields $\tau_{\mathrm{coll}} \propto E(t)^{0.24}$, for the same $\tau_0$, which assumes no time domain restrictions, as opposed to $\tau_{\mathrm{coll}}\propto t$. The parameters of the model, particularly the fractional energy gain $\alpha$, and the growth rate $s$, are very similar to those obtained with a mass ratio of 16. Nevertheless, the parameter study required to explore these dependencies, and to completely understand the microphysics underlying the parameter values, is beyond the scope of the present paper and will be tackled in future work. 

\section{Conclusions \& Discussion}

Ab-initio full PIC simulations have been presented for a two-dimensional relativistic collisionless shock propagating in an initially unmagnetized electron-ion plasma (mass ratio of 32). The shock structure and jump conditions confirm previous results obtained with a different PIC framework \citep{spitk08a,spitk08b}. Non-thermal particle acceleration is observed and occurs as particles are trapped and oscillate in the shock front, similarly to a Fermi acceleration process. Nevertheless, specific distinctions exist from the standard Fermi mechanism, namely the small number of scatterings and the continuous evolution of the shock structure where the particle accelerates. When gaining energy, the particle usually performs a rotation to the transverse direction, and thus remains in the shock region, being susceptible to further acceleration.

An important consequence of the overall acceleration efficiency ($\sim 10\%$ of energy carried by the most energetic particles) is the increased relevance of nonlinear effects for the shock structure and for the acceleration process. In fact, when the accelerated particles yield $\gtrsim 10\%$ of the plasma energy, their dynamics becomes relevant and influences the evolution of the overall system \citep{jones91}. One of these nonlinear effects is the dynamic pressure of the accelerated particles that slows down the unshocked plasma before it reaches the sharp shock transition. This is indicated in the simulation results as the shock widens and the magnetic and electric field layers extend to the upstream \citep{keshet08,medvedev08}. Furthermore, the inclusion of the trapped particles population can have implications on the jump conditions, and current models can be extended to incorporate these effects into the equation of state in the shock region. This generalization has been made for the non-relativistic and electrostatic shocks case \citep{fors70,sorasio06}.

The average energy growth of the particles can be reproduced by a multi-scattering acceleration mechanism with $\Delta E = \alpha E$, assuming an increase of the time between collisions. The long term acceleration implications of these effects cannot be inferred from our simulation spectra because of the exponential cut-off due to the finite simulated size that limits the injection of particles. Larger scale simulations with longer propagation distances and larger acceleration times will elucidate the properties of the particle spectrum at higher energies, thus allowing for a detailed identification of the mechanisms responsible for $\alpha \lesssim 1$, and for an increase of time between collisions. 

In summary, our results confirm the possibility of particle acceleration through a Fermi-like mechanism with a reduced number of energy gains, and generalized to reproduce the statistical data obtained with ab-initio full PIC simulations that self-consistently resolve the turbulent and non-linear evolution of the shock.

\vspace{0.2cm}

\small 
This work was partially supported by Funda\c c\~ao Calouste Gulbenkian and by Funda\c c\~ao para a Ci\^encia e Tecnologia under grants SFRH/BD/35749/2007 and PTDC/FIS/66823/2006 (Portugal). The simulations presented were produced using the IST Cluster (IST/Portugal) and the Dawson cluster (UCLA). The authors would like to thank Profs. A. Spitkovsky, S. Schwartz, and M. Medvedev for useful discussions.



\end{document}